\documentclass[aps,preprint,showpacs,pra]{revtex4}
\usepackage{epsfig,amssymb,amscd}
\begin{document}
\bibliographystyle{prsty}

\title{\bf The readout of the fullerene-based quantum computing by a scanning tunneling microscope}
\author{M. Feng$^{1,2}$, G.J. Dong$^{1}$ and B. Hu$^{1,3}$} 
\affiliation{$^{1}$ Department of Physics and Centre for Nonlinear Studies, and The Beijing-Hong Kong -Singapore 
Joint Centre for Nonlinear and Complex Systems (Hong Kong), Hong Kong Baptist University, Hong Kong, China \\
$^{2}$ Wuhan Institute of Physics and Mathematics, Chinese Academy of Sciences, Wuhan, 430071, China \\
$^{3}$ Department of Physics, University of Houston, Houston, Texas 77204-5005} 
\date{\today}

\begin{abstract}

We consider to detect the electron spin of a doped atom, i.e., a nitrogen or a phosphorus, caged in a fullerene by currently
available technique of the scanning tunneling microscope (STM), which actually corresponds to the readout of a qubit in the fullerene-based quantum 
computing. Under the conditions of polarized STM current and Coulomb blockade, we investigate the tunneling matrix elements involving the exchange 
coupling between the tunneling polarized electrons and the encapsulated polarized electron, and calculate the variation of the tunneling current 
with respect to different orientations of the encapsulated electron spin. 
The experimental feasibility of our scheme is discussed under the consideration of some imperfect factors.

\end{abstract}
\vskip 0.1cm
\maketitle

\section {introduction}

Quantum information processing (QIP) based on solid-state materials has attracted much attention over past few years due to the potential of 
scalability. Since the qubits are encoded in individual electron spins in most of the solid-state QIP schemes, how to efficiently detect a single 
electron spin has recently become a focus. Although there is no fundamental restriction for the single spin detection, it seems that, to do the qubit 
readout very well, we are expecting further development of detection technique with sufficient spatial and temporal resolution.
  
Some preliminary experiments have reached the single spin level so far. For example, Scanning- Tunneling- Microscope (STM) electron spin resonance 
(ESR) has demonstrated single molecule ESR spectroscopy of iron impurities in Silicon \cite {rao}, although theoretical work remains to clarify for 
the best description of the effects \cite {balatsky}. A very recent STM experiment under low temperature and with a high magnetic field has shown the spin flip 
spectra of a single manganese atom \cite {eigler}, and the magnetic resonance force microscope, which utilizes a cantilever driven by spins oscillating in 
resonance, has successfully detected a single surface electron spin \cite {rugar}. Moreover, it has been reported that 
micro-superconducting quantum interference devices are capable of distinguishing large spin difference ($\Delta m_S \sim 30$) \cite {pakes}.

There have been some theoretical proposals for single spin detection in the candidate systems of QIP \cite {kane,loss}. The present work will focus on the 
detection of a single electron spin of an encapsulated atom inside a fullerene (i.e., C$_{60}$), which corresponds to the readout of a qubit in performance of 
quantum computing based on the endohedral fullerenes $N @ C_{60}$ or $P @ C_{60}$ \cite {har,jason1}. The electron spin inside the fullerene plays the 
role of a qubit or an auxiliary qubit. As the electrons of the doped atoms cannot escape the $C_{60}$ cage while preserving their spin 
states, we must develope special proposals for the spin detection, different from in \cite {kane,loss}. In fact, there have been some schemes
in this respect, for example, by using a modified single molecule transistor (SMT) \cite {feng1}, by a nanomagnet molecule Fe$_{8}$ \cite {feng2}, and 
by a shuttling device \cite {jason2}. However, no relevant experiment has been achieved yet.

In the present paper, we investigate the possibility to detect a single electron spin inside the fullerene by STM. STM is a mature technique available
to manipulate single atoms with high precision at low temperature \cite {eigler2}. Although no reliable evidence has been shown to achieve a single spin 
detection, there have been some experiments \cite {eigler,durkan,paolo} and theoretical proposals \cite {bala1,bala2,bala3} regarding single spin problems with 
the STM. Our scheme is different from those proposals in following points: First, the electron spin to be detected, in our case, is assumed to be  
well polarized along an always-on magnetic field, while we don't know whether the spin is up- or down- polarized. This corresponds to the final 
result of a real quantum computing, i.e., the qubits returning to product states from entanglement and superposition after the elaborately designed 
logic gates have been carried out. So our detection is to know the orientation of the spin polarization. Second, as the tunneling through the air is 
different from that through a fullerene, our treatment is to separate the whole tunneling process into three consecutive steps. Third, due to the C$_{60}$ 
cage, the local electron spin (i.e., the spin to be detected inside the cage) is well protected. As a result, different from the case in 
\cite {bala1,bala2,bala3} with the electron spin on the surface, the main error in our case is from the vibration of the fullerene, instead of 
the spin scattering, due to the tunneling electrons.  
 
The conditions we employ in the treatment below include Coulomb blockade \cite {park} and the polarized current in STM \cite {heize}, both of which have 
been achieved experimentally.
The Coulomb blockade restricts the fullerene to be charged by no more than one electron, which means an electron from the STM tip could jump on the 
C$_{60}$ only after the previous electron sitting on the C$_{60}$ has jumped away. This has been demonstrated in a recent experiment \cite {park} for SMT. 
So we will suppose that this Coulomb blockade works throughout our scheme under a suitable bias voltage of the STM. The polarized current in STM implies 
that the electrons going out of the STM tip are well polarized, i.e., up or down-polarized. It is reported in \cite {heize} that the high-quality 
polarized current is already available in STM. So we suppose below that the tunneling 
electrons out of the STM tip are up-polarized. In the presence of a magnetic field, the tunneling electrons will couple to the caged electron by exchange 
interaction due to the spin degrees of freedom involved \cite {elste}. Moreover, the total electron spin of the doped atom N or P are $S=3/2$ with 
four Zeeman levels $|\pm 3/2\rangle$ and $|\pm 1/2\rangle$ \cite {jason1}. It has been demonstrated that quantum gating can be performed independently 
with electron spin states $|\pm 3/2\rangle$ or $|\pm 1/2\rangle$ \cite {feng3}. For simplicity, however, we will focus below on the discussion about the 
caged electron spins $|\pm 1/2\rangle$. The case regarding 
$|\pm 3/2\rangle$ could be obtained simply by enlarging the variation of the tunneling current by three times.

In the next section, we will explore the spin-dependent tunneling matrix elements. Then a specific calculation of the tunneling will be made in Sec III 
which also includes discussions about the experimental imperfection. The last section is for the conclusion. 

\section {the spin-dependent tunneling}

To clarify our description below, we simplify the system shown in Fig. 1 to be a doped fullerene sandwiched by two leads L and R, as shown in 
Fig. 2. So the Hamiltonian is,
\begin{equation}
H=\sum_{\lambda,k,l} \epsilon_{\lambda,k} c^{\dagger}_{\lambda,k,l}c_{\lambda,k,l} + H_{c}+ \sum_{\lambda,k,n,l} 
(t_{\lambda} c^{\dagger}_{\lambda,k,l} a_{n,l} + t^{*}_{\lambda} c_{\lambda,k,l} a^{\dagger}_{n,l}),
\end{equation}
where $c^{\dagger}_{\lambda,k,l}$ ($c_{\lambda,k,l}$) creates (annihilates) an electron in the lead $\lambda=L,R$, with $k$ being the momentum of 
the electron, $l=\pm 1$ for (up/down) spin polarization, and $\epsilon_{\lambda,k}$ are energies regarding Fermi energy. The operator $a^{\dagger}_{n,l}$ 
($a_{n,l}$) is related to the electron on the fullerene in the orbital $n$. $t_{\lambda}$ is the tunneling coefficient regarding $\lambda = $ L or R. 
$H_{c}$ accounts for 
the fullerene including the energy of itself and the Coulomb blockade term $(\sum_{k} a^{\dagger}_{k}a_{k})(\sum_{k} a^{\dagger}_{k}a_{k} - 1)$
as well as the exchange coupling $-J\hat{S} \cdot \hat{\sigma}$ with $\hat{\sigma}$ and $\hat{S}$ the spin degrees of freedom of the electrons outside 
and inside the fullerene, respectively. Since our interest is in the tunneling modification due to the electrons' exchange interaction, instead of 
the electro-phonon \cite {mil} or spin-phonon resonances, we may follow the idea in \cite {bala1,bala2,bala3} and focus on the spin dependent 
tunneling matrix elements. To this end, we separate the tunneling into three steps (See Fig. 2): The first step is from the left lead to the fullerene, 
the second one is for tunneling through the fullerene, and the third accounts for the tunneling from the fullerene to the right lead. 

Both the first and the third steps can be modeled as electronic tunneling through capacitors with $\Phi_{i}=W_{i} d_{i}$ where $i=1, 3$,
$\Phi_{i}$ is the work function, $d_{i}$ is the tunneling gap, and $W_{i}$ represents the constant regarding the free space permittivity, the dielectric 
constant and the area of the lead. By using WKB approximation, we could reach the spin-dependent tunneling matrix elements as below \cite {explain2}, 
\begin{equation}
\hat{T}_{1} \propto \exp \{-\sqrt {\frac {8m}{\hbar^{2}}} \int^{x_{1}}_{0} \sqrt {\Phi_{1} - J\hat{S}\cdot \hat{\sigma}} dx\},
\end{equation}  
and
\begin{equation}
\hat{T}_{3} \propto \exp \{-\sqrt {\frac {8m}{\hbar^{2}}} \int^{x_{3}}_{x_{2}} \sqrt {\Phi_{3} - J\hat{S}\cdot \hat{\sigma}} dx\},
\end{equation} 
where $J$ is the strength for the exchange coupling between the tunneling electrons and the caged electron due to the spin degrees of freedom of the 
electrons, which is proportional to the integral below
\begin{equation}
\int\int \Psi^{*}_{1}(r_{1})\Psi^{*}_{2}(r_{2}) \frac {e^{2}}{r_{12}} \Psi_{1}(r_{2})\Psi_{2}(r_{1}) dr_{1}dr_{2} 
\end{equation}
with $\Psi_{1}(r_{i})$ and $\Psi_{2}(r_{j})$ the wavefunctions of the electrons tunneling and caged at the position $r_{i}$ and $r_{j}$, respectively,
and $r_{12}=|r_{1}-r_{2}|$. For simplicity, we have omitted in Eqs. (2) and (3) the unimportant constants regarding the spin independent tunneling 
rate in the absence of $J$. Please note, we have also  made an assumption here that,
due to the Coulomb blockade and the low temperature limits, each electron from the left lead would jump to the ground state of $C_{60}^{-}$ and the 
tunneling amplitude is mainly dependent on the exchange interaction, instead of the electronic momenta. Since both spins of the tunneling and caged 
electrons are well polarized, we may simplify $J\hat{S}\cdot \hat{\sigma}$ to be $J\sigma_{z}$ with $J$ being 1 meV \cite {elste}. As a result, 
straightforward algebra for Eq. (2) yields corresponding tunneling rates,
\begin{eqnarray*}
T_{1} & \propto & \exp \{- \frac {2}{3}\sqrt {\frac {8mW_{1}}{\hbar^{2}}} [(x_{1}\pm \frac {J}{W_{1}})^{3/2}-(\pm \frac {J}{W_{1}})^{3/2}]\} \\
& \approx &  \exp \{- \frac {2}{3}\sqrt {\frac {8mW_{1}}{\hbar^{2}}} x_{1}^{3/2} (1\pm \frac {3J}{2W_{1}x_{1}})\},
\end{eqnarray*}
where the signs '$\pm$' correspond to the up/down polarization of the caged electron spin. We have approximately omitted in above equation the high-order 
terms regarding $J/W_{1}$ because they are comparatively negligible (i.e., $J/W_{1} \sim x_{1}/1000$ to the numbers shown later). Similarly, we have
\begin{eqnarray*}
T_{3} & \propto & \exp \{-\frac {2}{3}\sqrt {\frac {8mW_{3}}{\hbar^{2}}} [x_{3}^{3/2} (1\pm \frac {3J}{2W_{3}x_{3}})-x_{2}^{3/2} 
(1\pm \frac {3J}{2W_{3}x_{2}})]\} \\
& \approx & \exp \{-\frac {2}{3}\sqrt {\frac {8mW_{3}}{\hbar^{2}}} x_{3}^{3/2} (1\pm \frac {3J}{2W_{3}x_{3}})\}  
\exp \{\frac {2}{3}\sqrt {\frac {8mW_{3}}{\hbar^{2}}} x_{2}^{3/2} (1\pm \frac {3J}{2W_{3}x_{2}})\}. 
\end{eqnarray*}

The second step can be considered as a tunneling through a constant potential barrier. This is because the fullerene could be modeled as a spherical 
capacitor. Due to the equipotential surface of the spherical capacitor,
the work function $\Phi_{2}$ should be a constant, which yields the tunneling rate as
\begin{equation}
T_{2} \propto \exp \{-\sqrt {\frac {8md_{2}^{2}}{\hbar^{2}}(\Phi_{2} \pm J)} \}.
\end{equation}
Since $J$ is smaller than $\Phi_{i}$ by 1000 times, we may write the terms regarding $J$ as
\begin{equation}
\exp \{\pm D_{i}\}= \cosh (D_{i}) \pm \sinh (D_{i}) \approx 1\pm D_{i},
\end{equation}  
where $D_{1}=J\sqrt{8mx_{1}/(W_{1}\hbar^{2})}$ is from $T_{1}$, $D_{2}=J\sqrt{8mx_{3}/(W_{3}\hbar^{2})}$ and $D_{3}=J\sqrt{8mx_{2}/(W_{3}\hbar^{2})}$ 
are from $T_{3}$, and $D_{4}=\sqrt{2m}Jd_{2}/\sqrt{\Phi_{2}\hbar^{2}}$ is from $T_{2}$.

Since the tunneling current is proportional to the multiplication of the tunneling rates above, we may simply write 
the tunneling current as
\begin{equation}
I = I_{0} \prod_{i=1}^{4} (1\pm D_{i}),
\end{equation}
where
\begin{equation}
I_{0} \propto \exp \{-\sqrt{\frac {32mW_{1}}{9\hbar^{2}}}x_{1}^{3/2}\} \exp \{-\sqrt{\frac {32mW_{3}}{9\hbar^{2}}}(x_{3}^{3/2}-x_{2}^{3/2})\}
\exp\{-\sqrt{\frac {8m\Phi_{2}}{\hbar^{2}}}d_{2}\},
\end{equation}
is the spin independent tunneling current, which could be detected in the absence of the magnetic field. In the following calculation, we assume 
$I_{0}$ to be 1 nA \cite {park} if $d_{1}=0.23$ nm. 

\section{Discussion}

From \cite {park} we know that, $d_{2}=0.7$ nm is the size of the fullerene, and $d_{3}=0.27$ nm is due to van der Waals interaction between 
the fullerene and the substrate. We suppose that $\Phi_{1}(x)$ is 2 eV between $x=0$ and $x=x_{1}$, $\Phi_{2}=1$ eV is constant, and 
$\Phi_{3}(x)=1$ eV is from $x=x_{2}$ to $x_{3}$. Then we can obtain by direct calculation that $\Delta I_{\pm} = I_{\pm}-I_{0}= \pm 16$ pA, where 
$I_{\pm}$ corresponds, respectively, to $\pm$ in Eq. (7). The current of 16 pA is detectable with the present STM technology.  Slightly changing $d_{1}$, 
we may estimate $\Delta I= \Delta I_{+} - \Delta I_{-}$, which changes drastically with respect to $d_{1}$, as shown in Fig. 3. So it is evident that 
the difference between spin-up and spin-down of the caged electron is distinguishable from the tunneling current as long as $d_{1}$ is around 0.24 nm.

As mentioned above, because the local spin in our case is pretected by the fullerene cage, the main source of error is the vibration of the fullerene 
due to the tunneling electrons, which yields a shift $\delta=3$ pm equivalent to the energy variation 5 meV \cite {park}. Because this vibrational 
degree of freedom is far detuned from other characteristic frequencies, and also because there is no evidence of vibration-spin coupling, we did not 
consider this vibration in above treatment. The tunneling current, however, is very sensitive to the distance variation. So we have to strictly assess the 
influence due to the position shift of the fullerene on the tunneling current. To this end, we consider the tunneling gaps regarding the first and the third 
steps to be changed time-dependently, while the second step remains unchanged. So we have the modification of the tunneling rates,
\begin{equation}
T_{1} \propto \exp \{- \frac {2}{3}\sqrt {\frac {8mW_{1}}{\hbar^{2}}} x_{1}^{3/2} (1 + \frac {3\Delta}{2x_{1}} \pm \frac {3J}{2W_{1}x_{1}})\},
\end{equation} 
\begin{equation}
T_{3} \propto \exp \{-\frac {2}{3}\sqrt {\frac {8mW_{3}}{\hbar^{2}}} x_{3}^{3/2} (1\pm \frac {3J}{2W_{3}x_{3}})\} 
\exp \{\frac {2}{3}\sqrt {\frac {8mW_{3}}{\hbar^{2}}} x_{2}^{3/2} (1 -\frac {3\Delta}{2x_{2}} \pm \frac {3J}{2W_{3}x_{2}})\}.
\end{equation}
Suppose $\Delta = (\delta/2) \cos(\omega t)$ with $\omega = 1$ THz \cite {park}. Like in 
\cite {bala1}, we introduce the average current over the time $T$, i.e., $ \langle I \rangle = (1/N) \sum_{i=1}^{N} I (t_{i})$, where the sum from $i=1$ 
to N is over the number of the tunneling electrons in the time $T$, and $I (t_{i})$ is proportional to the tunneling rate for each electron.
So $ \langle I \rangle $ can be written as
\begin {equation}
\langle I \rangle = I_{0} \prod_{i=2,4}(1 \pm D_{i}) [1 \pm D_{1} + (\delta/2 x_{1}) \sqrt{\frac {8mW_{1}}{\hbar^{2}}} 
\int^{T}_{0} (1/T) \cos(\omega t) dt][1 \pm D_{3} 
\end {equation}
$$ - (\delta/2 x_{2})\sqrt{\frac {8mW_{3}}{\hbar^{2}}} \int^{T}_{0} (1/T) \cos(\omega t) dt]. $$
As the terms regarding $\delta/x_{1(2)}$ are comparable to $D_{1(3)}$, we have to eliminate the vibrational influence. To this end, we may detect the 
current during a time period $T= k\pi/\omega$ with $k=1, 2, \cdots$, in order to average out the terms regarding $\delta$. This would in principle make 
the fullerene vibration, due to the tunneling of the electrons, of negligible effect on the original current in our observation. However, due to
the intrinstically stocastic property of the electron tunneling, the efficiency of above trick depends on the statistical deviation $\sigma$ of a normal
distribution $(1/(\sqrt{2\pi}\sigma)\exp {[-(t-t_{0})^{2}/(2\sigma)^{2}]}$, where $t_{0}$ is the mean tunneling time of each electron we desire 
\cite {feng1}.  To carry out our scheme to the best, we require $\sigma$ to be as small as possible.   

Besides, there are other possible imperfection in implementing our scheme. In a real quantum computing, the qubits are 
not usually well polarized at the end of the computing operations. For example, the C$_{60}$-caged electron-spin would probably be in superposition 
$F|\uparrow\rangle + G |\downarrow\rangle$ ($F\gg G$ or $F\ll G$) due to some unpredictable error sources in quantum computing. The estimated current 
dispersion due to this imperfect polarization is \cite {bala1}, e.g., $F\gg G$, 
\begin{equation}
\frac {\sqrt{\overline{\langle \Delta I^{2}\rangle}}}{I_{0}} \approx \frac {2G}{\sqrt{\overline{N}}} \sum_{i=1}^{4} D_{i},
\end {equation}
in which $G$ is much smaller than 1 and $N$ is of the order of hundred in an observation during a period of $T= $100 ns. So  this current dispersion
is too small to affect our spin detection. In addition, the feedback effect of the tunneling electrons on the caged spin should also be addressed. This
estimate could be done by Fermi golden rule from the second-order perturbation \cite{bala1,bala2}, which yields the spin decay rate 
$1/\tau_{s}=|\sum_{i=1}^{4} D_{i}|^{2}/\tau_{e}$ with $1/\tau_{e}$ the tunneling rate of the electrons. As we suppose $I_{0}=1$ nA, implying 
$1/\tau_{e}=10^{10}$ Hz, we have $\tau_{s}\ge 10^{-6}$ sec, which is much longer than our detection time. So this spin decay is also negligible. 

Normally, the tunneling current in usual use of STM is from the order of pA to nA, which implies that the tunneling rate of the mobile electrons is from 
one electron every 100 ns to one electron every 100 ps. We have noticed that the observation of individual electron-jump is within the reach of the 
current technique 
by I-V characteristic and by dI/dV plot \cite {noguchi}, and the minimum current the STM is able to distinguish is of the order of 0.1 pA \cite {explain}. 

Recent experiment \cite {yama} has shown the availability of a controllable manipulation of a single C$_{60}$ by STM. Moreover, in the absence of
the tunneling of the electrons outside the fullerene, the decoherence time of 
the caged electron-spin is about 1 second at 7$^{o}$ K \cite {jason1}, much longer than our implementation time. So decoherence is neglected in 
our treatment above. While there is no experimental data for dephasing of the tunneling electrons. A recent work \cite {vn} has shown that 
the spin dephasing time of the bulk two-dimensional electron gas at low temperature could be 150 ns. In our case, we require the 
dephasing time of the tunneling electrons to be longer than 100 ns.  Considering the data in \cite {vn} and the difference between the bound and 
unbound electrons, our requirement should be satisfied at very low temperature.

Compared with a previous work \cite {feng1} using a modified SMT, the present scheme is, to some extent, similar, but much simpler.
In \cite {feng1}, spin flips of the mobile electrons by microwave pulses are necessary. Since the electron tunneling 
is intrinsically stocastic, the decoherence due to imperfect spin flip yields the main infidelity. In contrast, no spin flip is 
needed in the present scheme. So the dephasing of the mobile electron is not a serious problem in our scheme. 
Furthermore, a magnetic field gradient $\partial B/\partial z = 4 \times 10^{6}$ T/m is essential to \cite {feng1}.
While such a magnetic field being highly stable in time and very large and homogeneous in spatial gradient is still challenging with 
the present technique. In contrast, the constant magnetic field required in the present scheme is fully within the reach of the current technique.
Shortly speaking, the present scheme is better than in \cite {feng1}, and the advantages are from a combination of STM and SMT: STM makes sure that 
the tiny variation of the current can be distinguished, and the Coulomb blockade in SMT experiment guarantees that the
change of the tunneling current corresponds to the fullerene charged by only a single polarized electron.

\section{conclusion and acknowledgement}

In summary, we have explored the possibility to detect the electron spin of the doped atom in a fullerene cage by STM, assisted by the Coulomb 
blockade and polarized current. Different from previous studies, we separated the whole tunneling process into three parts. After presenting the 
spin dependent tunneling matrix elements, we have shown by specific calculation the feasibility 
of this detection with current STM  technique. The detrimental influences from some imperfect factors have also been considered in our treatment. 
Since the STM is widely employed in atomic and molecular control in various systems, we argue that the main idea of the present 
work, i.e., the use of the STM device along with the Coulomb blockade and the polarized current could be in principle applied to other candidate systems 
of QIP for detection of a single electron spin. 

This work is supported in part by grants from Hong Kong Research Grants Council (RGC) and the Hong Kong Baptist University Faculty Research Grants 
(FRG). MF also acknowledges thankfully the support from NNSFC No. 10474118.

\begin {center}{\bf The captions of the figures}\end{center}

Fig. 1  Schematic of the detection of a single caged electron spin by STM, where the black dots mean the electrons, and $\bf \sigma$ and $\bf S$ correspond 
to the spin degrees of freedom regarding the electrons tunneling and caged, respectively. The caged electron spin can be $\pm 1/2$ or $\pm 3/2$, which 
pair of spins employed in quantum computing has been known before the detection. 
So what we want to detect is whether the spin is up (i.e., 1/2 or 3/2) or down (i.e., -1/2 or -3/2). The gap between the fullerene and the substrate is 
due to van der Waals interaction \cite {park}.

Fig. 2 (a) Schematic of our treatment to model the spin-dependent tunneling by three steps, and (b) the work functions in different steps are assumed.

Fig. 3 The current difference $\Delta I= \Delta I_{+} - \Delta I_{-}$ with respect to $d_{1}$.


\begin{thebibliography}{27}
\bibitem {explain2} It is a common understanding from the standard text book of quantum mechanics that the WKB method yields the tunneling amplitude to be proportional to 
$\exp \{-\sqrt {\frac {8m}{\hbar^{2}}} \int \sqrt {V-E} dx\}$, where $V$ and $E$ are, respectively, the potential barrier and the energy of the tunneling 
particles. Due to spin degrees of freedom involved, the potential is spin dependent in our case. So like in \cite {bala1,bala2}, we replace $(V-E)$ by 
$(\Phi - J\hat{S}\cdot \hat{\sigma})$ with $\Phi$ and $J$ being work function and the strength of the exchange coupling, respectively. The difference in
expressions of the tunneling matrix elements in \cite {bala1,bala2} from ours here is due to the constant potential barrier supposed in their treatment.
\end{thebibliography}
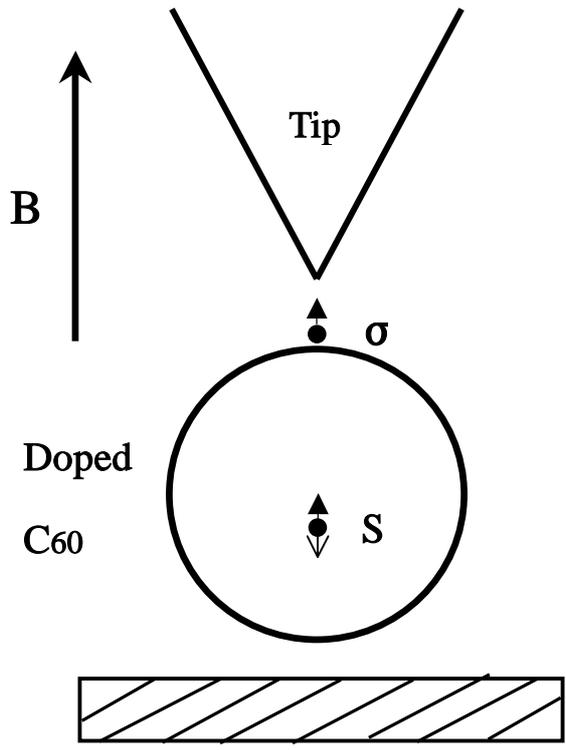
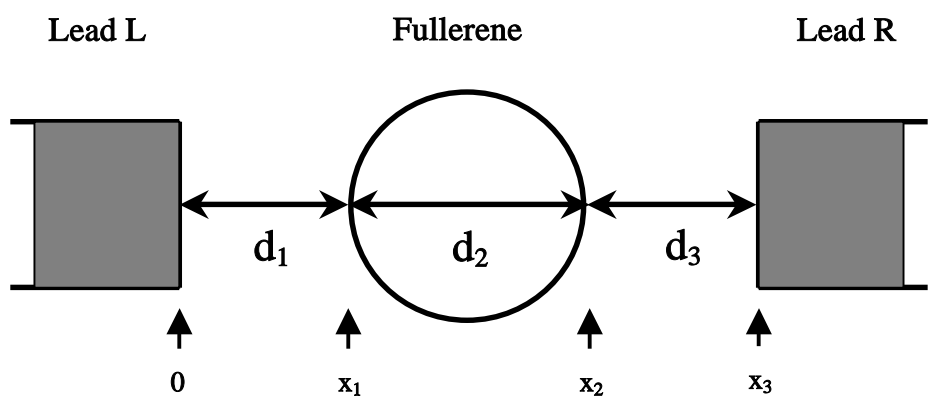
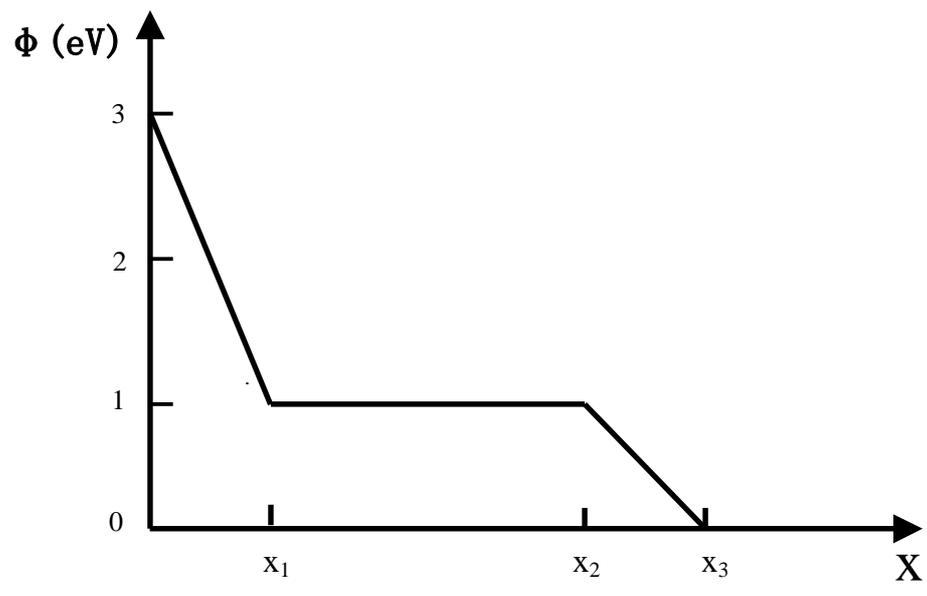
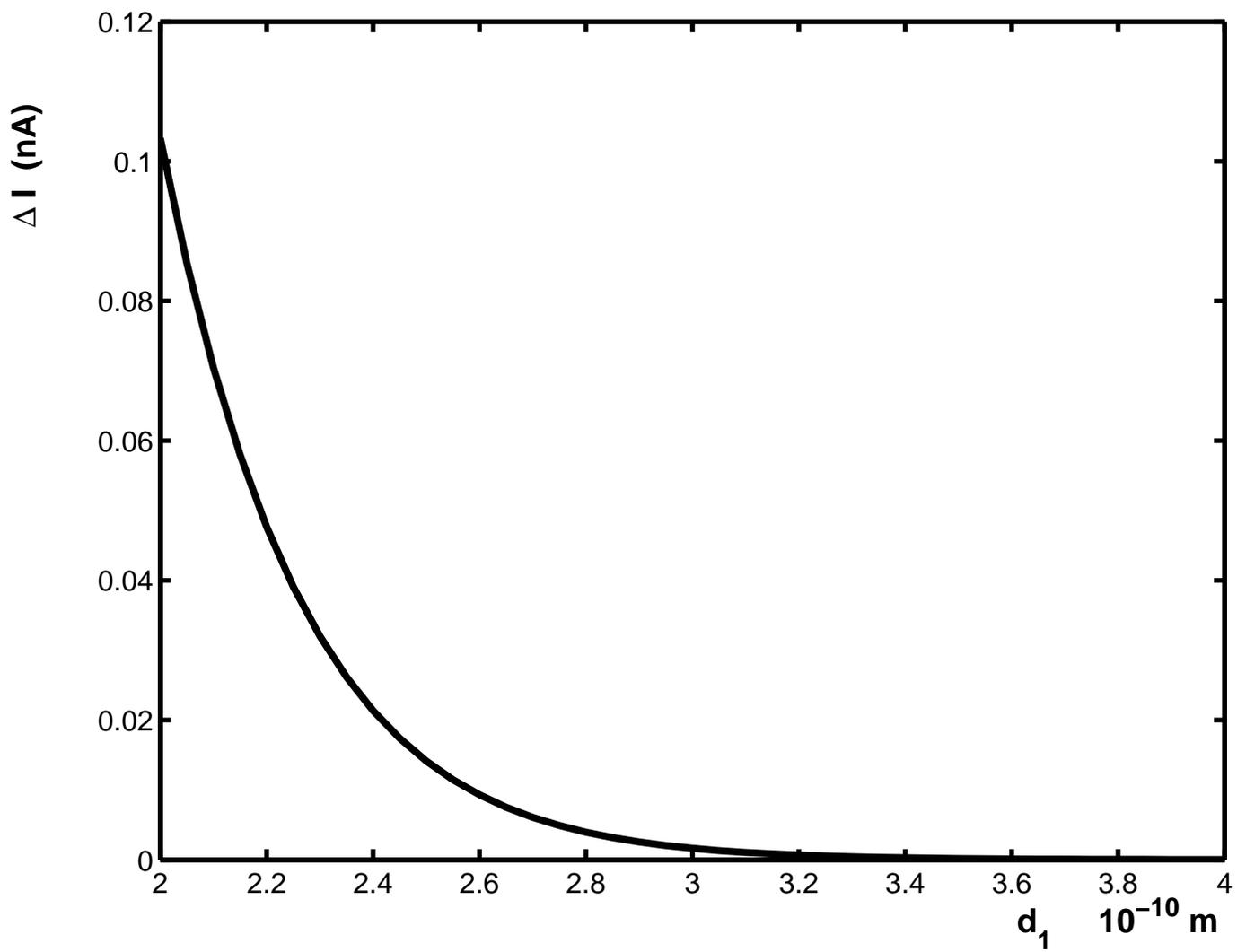
\end{document}